\documentclass[reprint,amsmath,amssymb,aps,superscriptaddress]{revtex4-1}

\usepackage{color}
\usepackage{amsmath}
\usepackage{graphicx}
\usepackage{dcolumn}
\usepackage{bm}
\usepackage{booktabs}
\usepackage{multirow} 
\usepackage{braket}
\usepackage[colorlinks]{hyperref}
\hypersetup{linkcolor = {blue},	citecolor = {blue},	urlcolor = {blue}}

\begin{document}
\preprint{APS/123-QED}

\title{Quantum Slide and NAND Tree on a Photonic Chip}

\author{Yao Wang}
\thanks{These authors contributed equally to this work}
\affiliation{Center for Integrated Quantum Information Technologies (IQIT), School of Physics and Astronomy and State Key Laboratory of Advanced Optical Communication Systems and Networks, Shanghai Jiao Tong University, Shanghai 200240, China}
\affiliation{CAS Center for Excellence and Synergetic Innovation Center in Quantum Information and Quantum Physics, University of Science and Technology of China, Hefei, Anhui 230026, China}

\author{Zi-Wei Cui}
\thanks{These authors contributed equally to this work}
\affiliation{Institute for Quantum Science and Engineering and Department of Physics, Southern University of Science and Technology, Shenzhen 518055, China}

\author{Yong-Heng Lu}
\affiliation{Center for Integrated Quantum Information Technologies (IQIT), School of Physics and Astronomy and State Key Laboratory of Advanced Optical Communication Systems and Networks, Shanghai Jiao Tong University, Shanghai 200240, China}
\affiliation{CAS Center for Excellence and Synergetic Innovation Center in Quantum Information and Quantum Physics, University of Science and Technology of China, Hefei, Anhui 230026, China}

\author{Xiao-Ming Zhang}
\affiliation{Department of Physics, City University of Hong Kong, Tat Chee Avenue, Kowloon, Hong Kong SAR, China}

\author{Jun Gao}
\affiliation{Center for Integrated Quantum Information Technologies (IQIT), School of Physics and Astronomy and State Key Laboratory of Advanced Optical Communication Systems and Networks, Shanghai Jiao Tong University, Shanghai 200240, China}
\affiliation{CAS Center for Excellence and Synergetic Innovation Center in Quantum Information and Quantum Physics, University of Science and Technology of China, Hefei, Anhui 230026, China}

\author{Yi-Jun Chang}
\affiliation{Center for Integrated Quantum Information Technologies (IQIT), School of Physics and Astronomy and State Key Laboratory of Advanced Optical Communication Systems and Networks, Shanghai Jiao Tong University, Shanghai 200240, China}
\affiliation{CAS Center for Excellence and Synergetic Innovation Center in Quantum Information and Quantum Physics, University of Science and Technology of China, Hefei, Anhui 230026, China}

\author{Man-Hong Yung}
\email{yung@sustech.edu.cn}
\affiliation{Institute for Quantum Science and Engineering and Department of Physics, Southern University of Science and Technology, Shenzhen 518055, China}
\affiliation{Shenzhen Key Laboratory of Quantum Science and Engineering, Southern University of Science and Technology, Shenzhen 518055, China}

\author{Xian-Min Jin}
\email{xianmin.jin@sjtu.edu.cn}
\affiliation{Center for Integrated Quantum Information Technologies (IQIT), School of Physics and Astronomy and State Key Laboratory of Advanced Optical Communication Systems and Networks, Shanghai Jiao Tong University, Shanghai 200240, China}
\affiliation{CAS Center for Excellence and Synergetic Innovation Center in Quantum Information and Quantum Physics, University of Science and Technology of China, Hefei, Anhui 230026, China}
\affiliation{Institute for Quantum Science and Engineering and Department of Physics, Southern University of Science and Technology, Shenzhen 518055, China}

\date{\today}

\maketitle

\textbf{In the age of post-Moore era, the next-generation computing model would be a hybrid architecture consisting of different physical components such as photonic chips. In 2008, it has been proposed that the solving of NAND-tree problem can be sped up by quantum walk. Such scheme is groundbreaking due to the universality of NAND gate. However, experimental demonstration has never been achieved so far, mostly due to the challenge in preparing the propagating initial state. Here we propose an alternative solution by including a structure called ``quantum slide", where a propagating Gaussian wave-packet can be generated deterministically along a properly-engineered chain. In this way, the optical computation can be achieved with ordinary laser light instead of single photon, and the output can be obtained by single-shot measurements instead of repeated quantum measurements. In our experimental demonstration, the optical NAND-tree is capable of solving computational problems with a total of four input bits, based on the femtosecond laser 3D direct-writing technique on a photonic chip. These results remove one main roadblock to photonic NAND-tree computation, and the construction of quantum slide may find other interesting applications in quantum information and quantum optics.}\\

\begin{figure*}
	\centering
	\includegraphics[width=1.9\columnwidth]{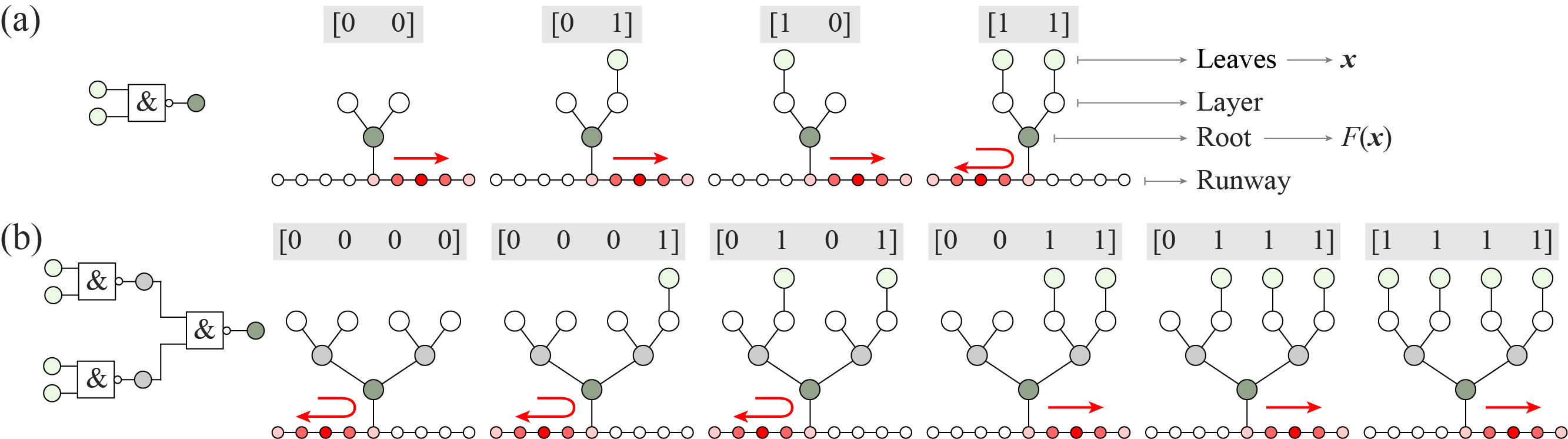}\\
	\caption{\textbf{Quantum NAND Tree.} The schematic of tree structure with one-layer branch (a), and two-layer branch (b). The site number in the last layer determines the number of input bits. The leaves on the last layer determine the input of the tree, if there is leaf on the last layer, then the input of this site is 1 otherwise is 0. It is a NAND gate for each site on the layer besides the last layer. The logical value of root presents the output result of the tree determining the evolution of wave-packet on the runway, the right-moving wave-packet will back if the output result of tree is 0, or go ahead if the result of tree is 1.}
	\label{f1}
\end{figure*}

Quantum walk, the quantum generalization of the classical random walk, is a natural platform for discovering exotic quantum phenomenon and developing quantum algorithms~\cite{QW_1, QW_1D, QW_2, QW_3, QW_co, QW_2D, photon_sim, FH}. Quantum walk based schemes have provided speedup for various problems of particular interest, including Boson sampling~\cite{BS_1, BS_2, BS_3, BS_4}, black-box problem~\cite{Box}, element distinctness~\cite{distinctness}, binary addition~\cite{QC_1}, factoring integers~\cite{QC_2}, and machine learning~\cite{ML}. It has also been shown that quantum walk is powerful enough to perform universal quantum computation~\cite{UCQW0, UCQW}.

A NAND tree is a binary tree of NAND gates containing a total of $N$ inputs but only one output, enabling the computation of arbitrary Boolean function of the form $F: \{0, 1\}^N\longrightarrow \{0,1\}$. 
While there are many classical algorithms for the NAND tree problem~\cite{Prob1, Prob2}, this task can be mapped to the quantum walk to speed up the computation. In 2008, Farhi \textit{et al.} proposed a continuous quantum walk based protocol for balanced NAND tree for the first time~\cite{farhi2007}. Inspired by it, several discrete quantum walk schemes have been developed~\cite{discrete, discrete_NAND}. The schemes have also been generalized to the unbalanced NAND formula~\cite{anyNAND, mm, anyANDOR}, which can in principle represent arbitrary Boolean functions~\cite{games}. 

The original scheme of NAND tree in Ref.\cite{farhi2007} is based on a continuous quantum walk on a graph. As shown in Fig.~\ref{f1}, the graph contains a chain of uniformly-coupled sites, called ``runway", a the quantum NAND tree whose sites are connected in the structure as a binary tree, and a set of input nodes. Suppose the graph is represented by the adjacent matrix $G$, the time-independent Hamiltonian of the entire system is just $H=J\cdot G$, where $J$ is the coupling constant. The input $\bm{x}$ is encoded by the on/off coupling of the input nodes and the leafs on top of the tree. For an incident wave-packet with zero energy in the run way~\cite{farhi2007}, if the computation outcome is $F(\bm{x})=1$, the wave-packet will pass through the quantum NAND-tree; otherwise the wave-packet with be reflected when $F(\bm{x})=0$. Therefore, by measuring whether the wave-packet has passed through or been reflected, the binary computation outcome can be obtained. 

The original protocol~\cite{farhi2007} of quantum NAND tree is elegant and simple, and there is also proposal of realizing it on the molecular platform~\cite{molecular_NAND} being developed recently. However, its experimental realization remains challenging and, to our best knowledge, has not been successfully demonstrated yet. The major difficulty lies in the preparation of the input state, where a truncated plane-wave with sharp boundaries was required.  

In this work, we propose an alternative approach by attaching a ``quantum slide" (QS) -- a chain with a parabolic shape in terms of the site-to-site coupling profile, to the runway. Consequently, a single photon (or even a light pulse) injected into a single site located at the edge of QS can naturally evolve through the quantum walk as a Gaussian wave-packet along the runway.

We experimentally demonstrate this new approach on a photonic chip, where the associated graph structure is fabricated with the femto-second laser direct-writing technique~\cite{fabri_1, fabri_2, fabri_3, PIT_Gap}. Our structure contains multi-waveguides with the number beyond 60, each of which corresponds to a node of the quantum walk. Instead of single photons, a light pulse is employed for preparing the Gaussian wave-packet, which allows us to efficiently obtain the computational results by single-shot measurements. We first verity that the wave-packet can be generated successfully by the QS, and transferred to the runway smoothly. Then, for both $N=2$ and $N=4$ cases, we measure the distribution of the photon intensity versus the evolution distance in the waveguide to verify the NAND tree logic. 

We first show how the Gaussian wave-packet can be used for the NAND-tree logic computation, and how QS can be employed for generating the Gaussian wave-packet. For a runway with a total of $L_{\text{rw}}$ sites labeled by $s$, the right-moving Gaussian wave-packet is described by
\begin{equation} 
|\psi_{\text{gs}}\rangle = A \sum_{s=1}^{L_{\text{rw}}} e^{-\frac{(s-\mu)^{2}}{4\sigma^{2}}} e^{-is\frac{\pi}{2}}|s\rangle,
\label{eq:wp}
\end{equation} 
where $\mu< L_{\text{rw}}/2$ is the center of the wave-packet, $A$ is the normalized coefficient $A=({1}/{2\pi\sigma})^{\frac{1}{4}}$, and $\sigma$ is the width of the wave-packet. Note that the NAND tree is ``planted" at the middle site, $s_{\text{mid}}=(L_{\text{rw}}+1)/2$, of the runway.

Eq.~\eqref{eq:wp} can be regarded as the superposition of a set of truncated plane-wave whose eigenenergy are close to zero. Therefore, its evolution is similar to the ideal plane-wave with zeros energy as proposed in~\cite{farhi2007}, and one can also obtain the computation outcome, $F(\bm{x})$, by measuring whether the wave-packet has passed through the NAND-tree. More specifically, we define $P_{+} = \sum _{s>s_{\rm{mid}}} \ket{s}\bra{s}$ as the projection on all the sites at the left of the middle site, and assume $\sigma$ increases quadratically with $N$, i.e. $\sigma=\gamma \sqrt{N}$ where $\gamma$ is a constant. The expectation value of $P_+$ after the wave-packet passed through or was reflected by the quantum the NAND-tree satisfies (see Supplemental Material)
\begin{equation} 
\langle P_{+}\rangle = F(\bm{x}) + O\left(\gamma^{-1/4}\right).
\end{equation}
We also find that to ensure the good performance, $L_{\text{rw}}$ should be proportional to $\sigma$. In particular, we numerically find that the optimum length is about $L_{\text{rw}}=6\sigma$. 

The Gaussian wave-packet in Eq.~\eqref{eq:wp} can be easily generated with a QS. For QS with $L_{\text{qs}}$ sites, the coupling strength between the $r$th and $(r+1)$th site of the QS is set as $J_{r}=J \sqrt{r(2L_{\text{qs}}-2-r)}/(L_{\text{qs}}-1)$, where $J$ is the coupling strength at the runway. This type of structure is first introduced for perfect state transfer purpose~\cite{PST}. Through quantum walk, a single excitation at site $r=1$ can generate a stable Gaussian wave-packet with high fidelity, and transfer it to the runway. The width of generated Gaussian wave-packet is related to the site number of QS as shown in Fig.~\ref{f2}(a), more details about the theoretical analysis and be found in Supplementary Materials.

\begin{figure*}
	\centering
	\includegraphics[width=1.8\columnwidth]{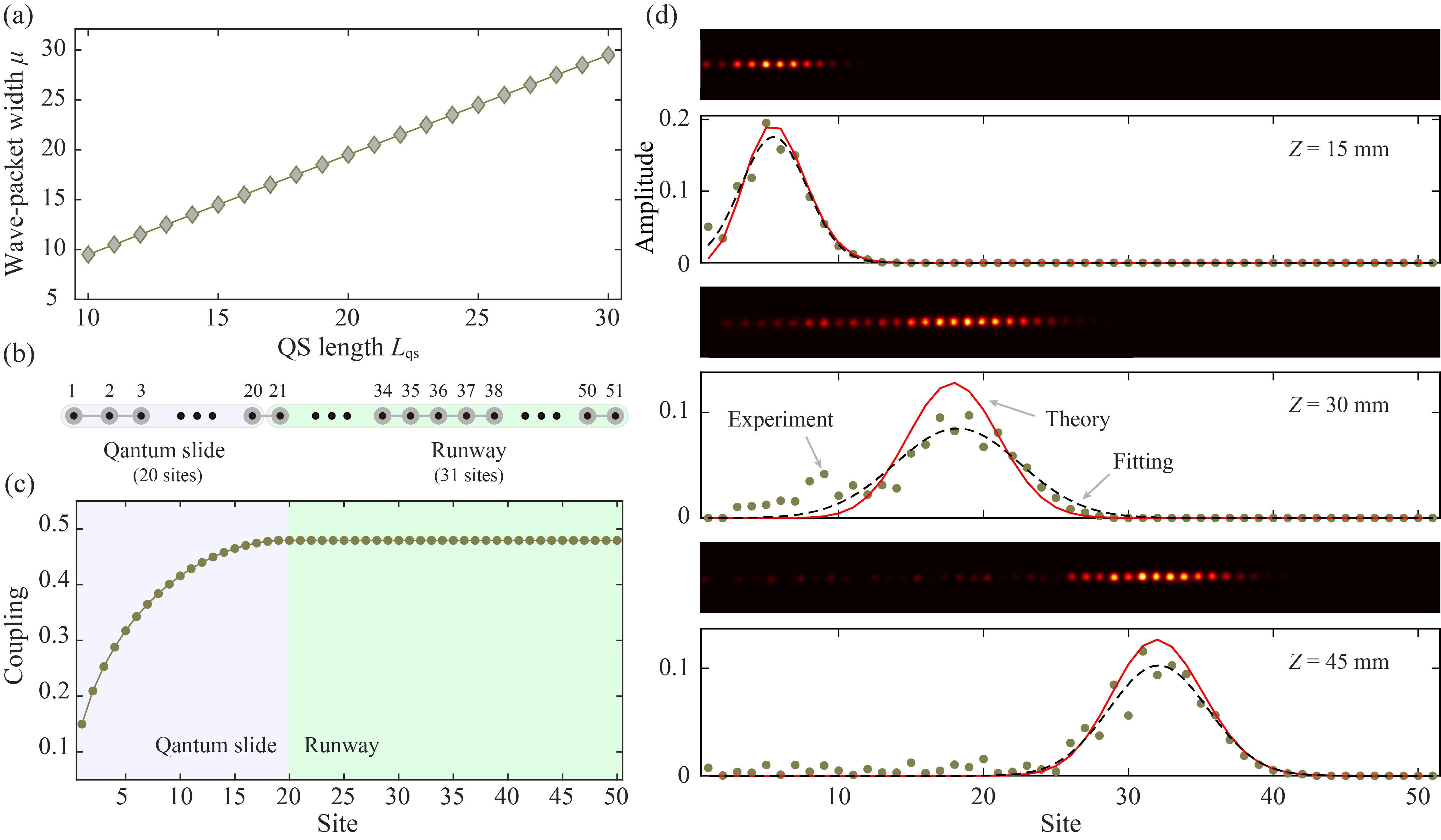}\\
	\caption{\textbf{Quantum slide.} \textbf{(a)} The relationship between the wave-packet width $\mu$ and the length of quantum slide $L_{\text{qs}}$. \textbf{(b)} The sketch of quantum slide. 20 sites is designed for the QS, the wave-packet obtained by the QS process transmits to the runway consisting of 31 sites. The sites in QS are labeled from 1 to 20 while the sites in runway are labeled from 21 to 51. \textbf{(c)} The distribution of the coupling strengths. \textbf{(d)} Measured output distribution of photons with different evolution distance in the QS process. The read lines are the theoretical results, the green points present the measured values, and the black dash lines are the fitting result of the experimental values.}
	\label{f2}
\end{figure*}

\begin{figure*}
	\centering
	\includegraphics[width=1.95\columnwidth]{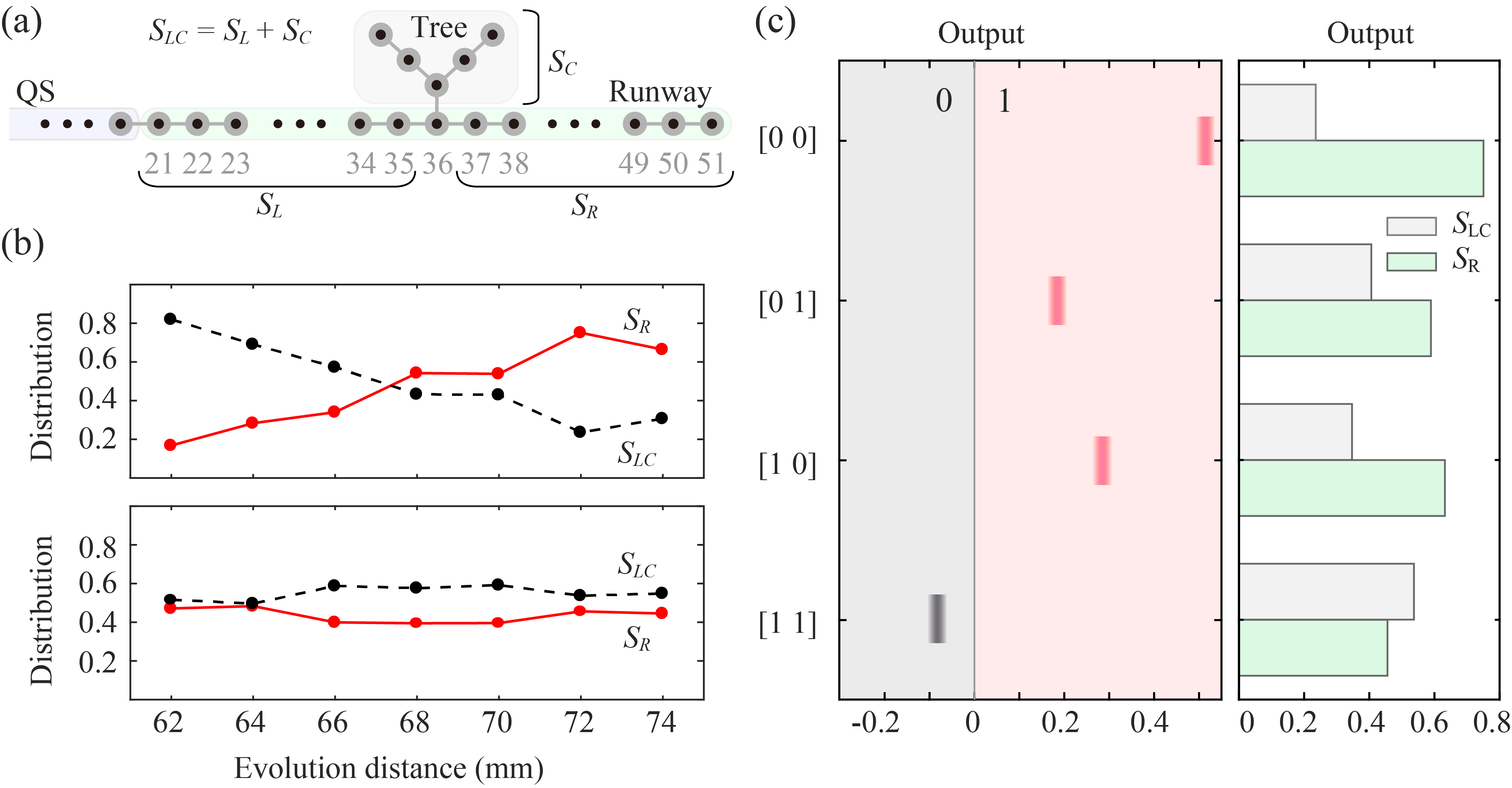}\\
	\caption{\textbf{Measured result of one-layer NAND Tree.} \textbf{(a)} The sketch of one-layer NAND Tree structure in experiment. $S_{\rm L}$ is the distribution amplitude summation of sites 21-35, $S_R$ presents the distribution amplitude summation of sites 37-51, $S_{\rm C}$ is the distribution amplitude summation of sites in NAND tree structure, and $S_{\rm LC}=S_{\rm L}+S_{\rm C}$. \textbf{(b)} The evolution result of $S_{\rm R}$ and $S_{\rm LC}$. For the case of input [0 0], the $S_{\rm R}$ go larger than $S_{\rm LC}$ with the increase of evolution distance, presenting the wave-packet go pass the site connecting the tree, while the result is contrary for the case of input [1 1]. \textbf{(c)} The measured result of the NAND tree.}
	\label{f3}
\end{figure*}

\begin{figure*}
	\centering
	\includegraphics[width=1.95\columnwidth]{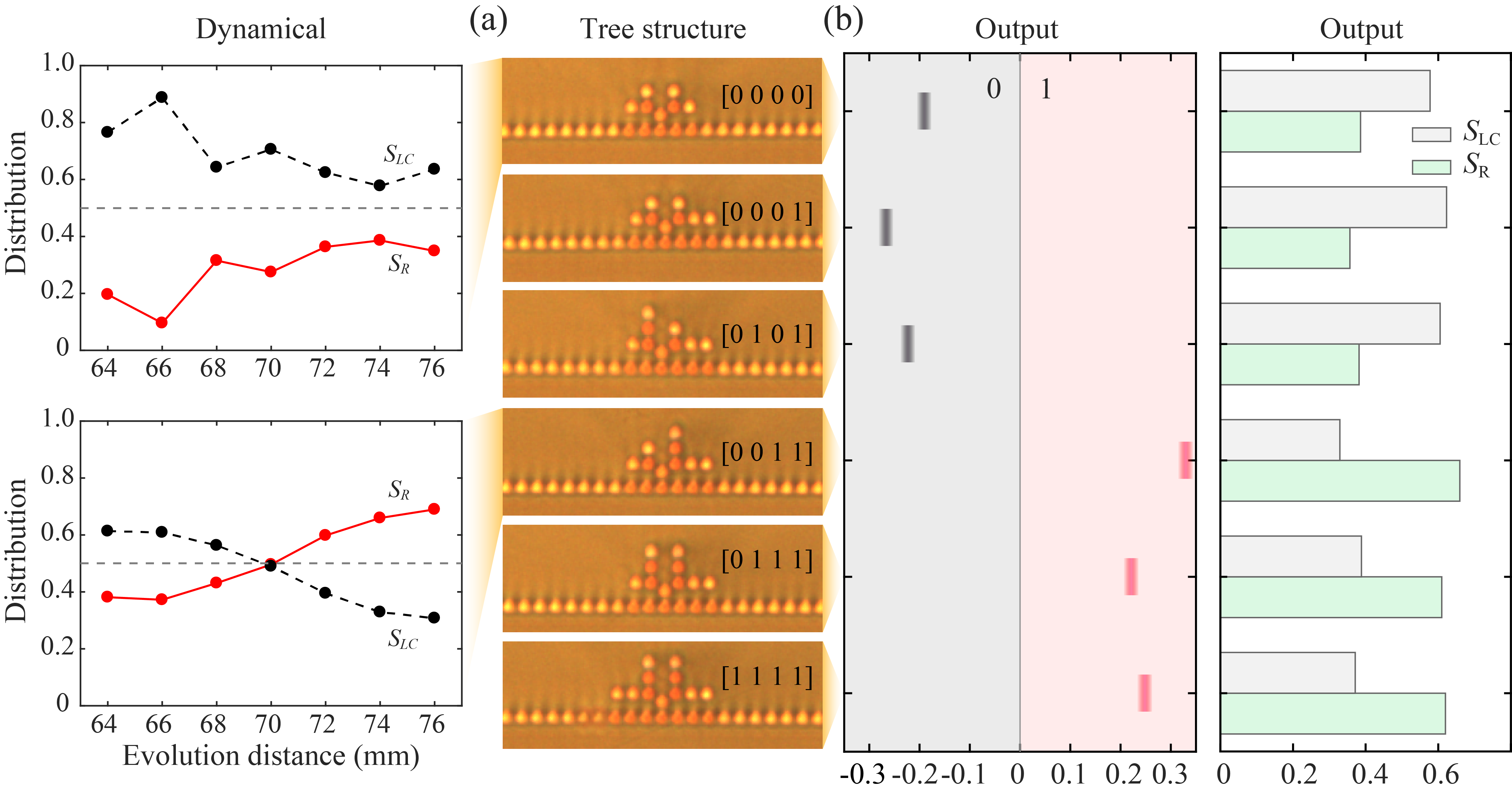}\\
	\caption{\textbf{Measured result of two-layer NAND Tree.} \textbf{(a)} The cross section micrograms of the NAND Tree. There are six types logical input for two-layer NAND tree. The inserts show the dynamical result for the cases of input [0 0 0 0] and [0 0 1 1] respectively. For the case of input [0 1 0 1], the $S_{\rm R}$ go larger than $S_{\rm LC}$ with the increase of evolution distance, presenting the wave-packet go pass the site connecting the tree, while the result is contrary for the case of input [0 0 0 0]. \textbf{(b)} The result of the NAND tree. The four-bit input quantum NAND algorithm is well realized and the visibility of output result is stable.}
	\label{f4}
\end{figure*}

As shown in Fig.~\ref{f2}(b-c), we set $L_{\text{qs}}=20$ and $L_{\text{rw}}=31$ respectively in our experiment. For clarity, we relabel the $s$th site of the runway as the $(r+20)$th site. By coupling the last site in the quantum slide ($r=20$) and the first site of the runway ($r=21$) with coupling strength $J$, the generated Gaussian wave-packet can be transferred to the runway smoothly. The structure accommodating the quantum walk experiment is fabricated in borosilicate glass using femtosecond laser direct writing technique, and the coupling strength is experimentally measured to be $J= 0.48$ mm$^{-1}$.

In Fig.~\ref{f2}(d), we show the measured photon distribution at evolution distance of 15, 30, and 45 mm. The results imply that the wave-packet is successfully generated and transferred to the runway as expected, and the wave-packet finally becomes stable and has a constant velocity in the runway. The measured velocity $v=0.925\pm0.027$ site/mm$^{-1}$ agrees well with the simulated result of  $0.941$ site/mm$^{-1}$, see more discussion in the Supplementary Materials.

The tree structure of two-bit input NAND logical algorithm contains one-layer branch, as shown in Fig.~\ref{f3}(a), and the root site can exchange the photon with middle site of runway labeled $r=36$. The branch sites bridge the two sites of leaves with the root site, and the leaf site determines the input as 1 or 0. We fabricate four lattices with same parameters but different inputs [0 1], [0 0], [1 0] and [1 1]. The photons with wavelength of $810$nm are transformed to pure horizontal polarization and then injected into the $r=1$ site by a 30X objective lens. The output photon distribution is measured using a 10X objective lens and a CCD camera [see Methods for details].

As discussed above, the NAND tree structure determines whether the wave-packet will pass the site $r=36$ connecting NAND tree. To quantify the result, we denote the sum of distribution probability over sites $21\leqslant r\leqslant 35$ as $S_{\rm L}$ and $S_{\rm R}$ as the sum of distribution probability over the sites $37\leqslant r \leqslant51$. In theoretical principle, we just need to simply compare the values of $S_{\rm L}$ and $S_{\rm R}$ to find whether the wave-packet is reflected or transmitted by the tree. However, considering the photons will also occupy the sites in the tree when they are reflected, which is not presented in the theoretical analysis. We further set $S_{\rm C}$ as the sum of distribution probability over site in the NAND tree. According to the theory, the output result of the NAND tree logic value is 1 if $S_{\rm R}>S_{\rm LC}$, where $S_{\rm LC}=S_{\rm L}+S_{\rm C}$, otherwise the output logic result is 0.

According the measured output distributions, we recognize the intensity of each site (see Methods for details) and normalize the distribution with $S_{\rm R}+S_{\rm LC}=1$. Taking the input [0 0] and [1 1] for example, we show the dynamical $S_R$ and $S_{\rm LC}$ in Fig.~\ref{f3}(b), the $S_{\rm R}$ finally becomes larger than $S_{\rm LC}$, presenting that the wave-packet goes pass to the right side of the runway, the corresponding result of quantum NAND tree is 1. It is contrary for the case of input [0 0] implying the logical result of 0. We define $L_{\rm out}=S_{\rm R}-S_{\rm LC}$ as the quantum NAND Tree logical output, the logical value is 0 if $L_{\rm out}<0$, otherwise, it is 1. The measured NAND Tree logical output and detailed values of $S_{\rm R}$ and $S_{\rm LC}$ of all the cases at evolution distance of 72 mm are shown in Fig.~\ref{f3}(c). Our experimental result imply that the fabricated tree structure successfully realize the quantum NAND algorithm.

For two-layer tree structure, there are four-bit input and the site number of NAND tree varies from 7 to 11. As shown in Fig.~\ref{f4}, there are six types logical operation and we fabricate all of these types in photonic chip. Taking the cases of input [0 0 0 0] and [0 1 0 1], the dynamical results of $S_{\rm R}$ and $S_{\rm LC}$ show the successful observation of logical output, implying that the evolution distance of 72 mm is the appropriate for evaluating the logical output of quantum NAND tree. We measure the outgoing photon distributions at evolution distance of 72 mm and analyze the result of $S_{\rm R}$, $S_{\rm LC}$ and $L_{\rm out}$ using the same methods mentioned above. The results shown in Fig.~\ref{f4} implies that the four-bit input quantum NAND algorithm is well realized and the visibility of output result is stable. Different from the one-layer NAND Tree, the two-layer NAND tree structure is able to well control the passing or blocking on the wave-packet in the runway in experiment. The reason behind may be that there are enough site number to influence the Hamiltonian of the whole lattice.

In conclusion, we present the first realization of quantum NAND Tree logic on integrated photonic chip. We theoretically propose and experimentally realize a way of preparing Gaussian wave-packet with Quantum Slide. Based on it, the NAND tree logic is successfully demonstrated for the structure with up to $N=4$ input.  The our work is based on the photonic chip with three-dimensional femtosecond laser direct writing technique. Compared to the molecular system~\cite{molecular_NAND}, such platform has great advanteges in the integration and scalability. Moreover, the balanced tree structure in this work can be easily generalized to the unbalanced NAND formula, which can be applied to the two-player games problems~\cite{games}. Finally, due to the universality of NAND gates, it is possible to generalize quantum NAND gate to represent arbitrary Boolean functions.\\

The authors thank Jian-Wei Pan for helpful discussions.  This research is supported by the Key-Area Research and Development Program of Guangdong Province (2018B030326001), 
the National Key R\&D Program of China (2019YFA0308700, 2017YFA0303700), 
the National Natural Science Foundation of China (61734005, 11761141014, 11690033,11875160, U1801661), 
the Science and Technology Commission of Shanghai Municipality (17JC1400403), 
the Shanghai Municipal Education Commission (2017-01-07-00-02-E00049),
the Natural Science Foundation of Guangdong Province (2017B030308003), 
the Guangdong Innovative and Entrepreneurial Research Team Program (2016ZT06D348), 
the Science,Technology and Innovation Commission of Shenzhen Municipality (JCYJ20170412152620376, JCYJ20170817105046702, KYTDPT20181011104202253), 
the Economy,Trade and Information Commission of Shenzhen Municipality (201901161512), 
X.-M.J. acknowledges additional support from a Shanghai talent program.\\

%

\subsection*{Methods}
{\bf Fabrication and measurement of the NAND Tree structure on a photonic chip:} We design the NAND Tree lattice structure according to the characterized relationship between the coupling coefficients and the separation of adjacent waveguides. The lattices are written into borosilicate glass substrate (refractive index $n_0=1.514$) with femtosecond laser (power 10W, wavelength 1026nm, SHG wavelength 513nm, pulse duration 290fs, repetition rate 1MHz). We reshape the focal volume of the beam with a cylindrical lens, and then focus the beam inside the borosilicate substrate with a 50X objective lens (NA=0.55),  A high-precision three-axis translation stage is in charge of moving the photonic chip during fabrication with a constant velocity of 10mm/s. 

The waveguides fabricated are round and the diameter is 3.75 $\mu$m. The separations between adjacent waveguides are determined by the coupling strength according to the characterized relationship between the coupling coefficients and the separation of adjacent waveguides.

In the experiment, the photons with wavelength of $810$nm are transformed to pure horizontal polarization by a quarter-wave plate, a half-wave plate and a polarized beam splitter. Then, we inject the photons into the lattice in the photonic chip using a 30X objective lens. After a total propagation distance through the lattice structures, the outgoing distributions photons are observed using a 10X microscope objective lens and the CCD camera.\\

{\bf Obtaining intensity distribution from CCD camera:} The intensity distribution of output photons is recorded by CCD camera with format of image and data. Part of images are shown in Supplementary Materials. We analyze the data to obtain the light intensity of each site with the position information which is determined before performing fabrication. The spot size is determined with same ratio of edge ($1/e$) to center value under the Gaussian fitting. Since all the unwanted output pattern of the scattered light is much weaker than the edge intensity, such that the intensity distribution can be obtained free of background noises.

\clearpage


\section*{Supplementary Materials}

\subsection{Theoretical analysis: quantum NAND tree}
In this section, the Quantum NAND tree problem will be theoretically analyzed in details. Followed by Farhi's work~\cite{farhi2007}, we will show that with Gaussian wave-packet, the algorithm can be realized with an arbitrary small error (Eq.~2 in the main text).

\subsubsection{The Structure of the Quantum NAND Tree Hamiltonian}
For simplicity, we relabel the site $s_{\rm{mid}}$ which attach to the root site of the quantum NAND tree as $r=0$, and let $r$ increases from left to right. 
We denote $H$ as the Hamiltonian of the system. For all sites $r\neq0$, we have 
\begin{equation}
	H\ket{r} = -\ket{r+1}-\ket{r-1},
\end{equation}
where $\ket{r}$ denotes excitation at site $r$. We consider the finite case, that is, the length of the runway $L_\text{rw}$ is finite. For simplicity, we denote $L_{\text{rw}}$ as $n$. Supposing matrix $M_n$ is an $n\times n$ matrix, in which all elements in vice diagonal are 1 and others are 0. $P_n$ is the eigen determinant of $M_{n}$ as

\begin{equation}
	P_{n}=\det(\epsilon I - M_n),
\end{equation}
where $\epsilon$ is the eigen value of $M_n$. It is easy to find that $P_n$ satisfies the second order liner recurrence relation:
\begin{equation}
	P_{n+2} = \epsilon P_{n+1} - P_{n}.
\end{equation}
By solving the recursion equation, we can obtain
\begin{equation}
	\epsilon = -2\cos{\frac{j\pi}{n+1}}.
\end{equation}
When $n \rightarrow \infty$, $\frac{j\pi}{n+1}$ could run over all values from 0 to $\pi$. In this way, we obtain the eigen energies of the Hamiltonian as $\epsilon = -2\cos{\theta} = E(\theta)$, and non-normalized eigen states are $e^{ir\theta}$ and $e^{-ir\theta}$, corresponding to the wave-packets moving in different directions. 
We can normalize the eigen state by
\begin{equation}
	\braket{r|E}=\left\{
	\begin{array}{lcl}
		e^{ir\theta}+R(E)e^{-ir\theta} & & {r\leq 0}\\
		T(E)e^{ir\theta} & & {r\geq 0}.\\
	\end{array} \right.\label{state}
\end{equation}

Now we consider the case at $r=0$, where the Hamiltonian is 
\begin{equation}
	H\ket{r=0}=-\ket{r=-1}-\ket{r=1}-\ket{\text{root}}.
	\label{r0}
\end{equation}
Here $\ket{\text{root}}$ is the excitation at the root site of the quantum NAND tree. Firstly, we can get a simple relationship as
\begin{equation}
	1+R(E)=T(E)\label{re}.
\end{equation}
Then taking inner product on Eq.\ref{r0} by $\ket{E}$, we have
\begin{equation}
	\braket{r=0|H|E}=-\braket{r=-1|E}-\braket{r=1|E}-\braket{\text{root}|E},    
\end{equation}
and using the fact that the eigen energy of the system is $H\ket{E} = -2\cos{\theta}\ket{E}$, we find
\begin{equation}
	-2\cos{\theta}\braket{r=0|E}=-e^{-i\theta}-R(E)e^{i\theta}-T(E)e^{i\theta} -\braket{\text{root}|E}.
	\label{s7}
\end{equation}
Inserting $\braket{r=0|E} = T(E)$ into Eq.~\ref{s7}, we can get
\begin{equation}
	2T(E)\cos{\theta} = -2i\sin{\theta} + 2T(E) e^{i\theta}+\braket{\text{root}|E}.
\end{equation}
Therefore, the transmit coefficient $T(E)$ is obtained as
\begin{equation}
	T(E)=\frac{2 i \sin \theta}{2i\sin\theta + \frac{\braket{\text{root}|E}}{\braket{r=0|E}}}.
\end{equation}

The transmit coefficient $T(E)$ depends on two parameters: the energy parameter $\theta$, and the fraction $y(E) = \frac{\braket{\text{root}|E}}{\braket{r=0|E}}$. For the case of $\theta = 0$, $T(E)$ will be 0 when the $y(E)$ is infinity, implying that the wave-packet is fully reflected, and the wave-packet will be fully transmitted when the $y(E)=0$. 

\begin{figure}[h]
	\centering
	\includegraphics[scale = 0.5]{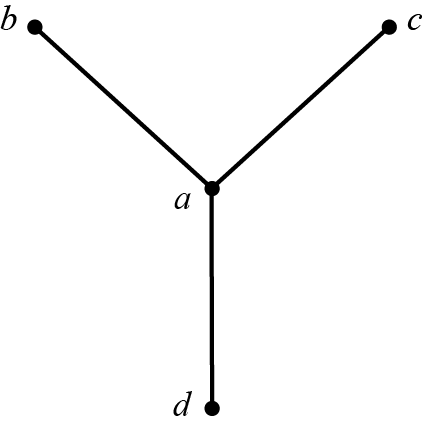}
	\caption{\textbf{One branch of the tree.} The fraction $Y=a/d$ is determined by inputs $Y_1=b/a$ and $Y_2=c/a$, and is the input of the lower layer branch. The behavior at $E=0$ corresponds to the NAND gate.}
	\label{fig_onebranch}
\end{figure}

In order to figure out the fraction $y(E)$, we firstly study the property of one branch in the tree, then generate to the whole part. As shown in Fig.~\ref{fig_onebranch}, we set $\ket{a}$ as the state that only site $a$ is $1$ (input nodes connected to the leafs of the tree) and other sites are 0 (input nodes disconnected to the leafs of the tree), and set lower letter $a$, $b$, etc as the amplitudes of $\ket{E}$ at the corresponding nodes $\ket{a}$, $\ket{b}$, etc. Setting $Y_1 = \frac{b}{a}$, $Y_2 = \frac{c}{a}$, $Y = \frac{a}{d}$, and applying Hamiltonian to site $a$, we have
\begin{equation}
	H\ket{a} = - \ket{b} - \ket{c} - \ket{d},
\end{equation}
\begin{equation}
	E = -\frac{b}{a}-\frac{c}{a}-\frac{d}{a},
\end{equation}
and
\begin{equation}
	Y = -\frac{1}{E+Y_1+Y_2}.
\end{equation}
If the inputs are restricted to either 0 or $\infty$, the branch will give an output $Y=-\infty$ when both $Y_1$ and $Y_2$ are $0$, otherwise $Y$ will give the output as 0, which corresponds to the NAND gate if we consider $Y=-\infty$ as bit 0 and $Y=0$ as bit 1.

\begin{figure*}[h]
	\centering
	\includegraphics[scale = 0.45]{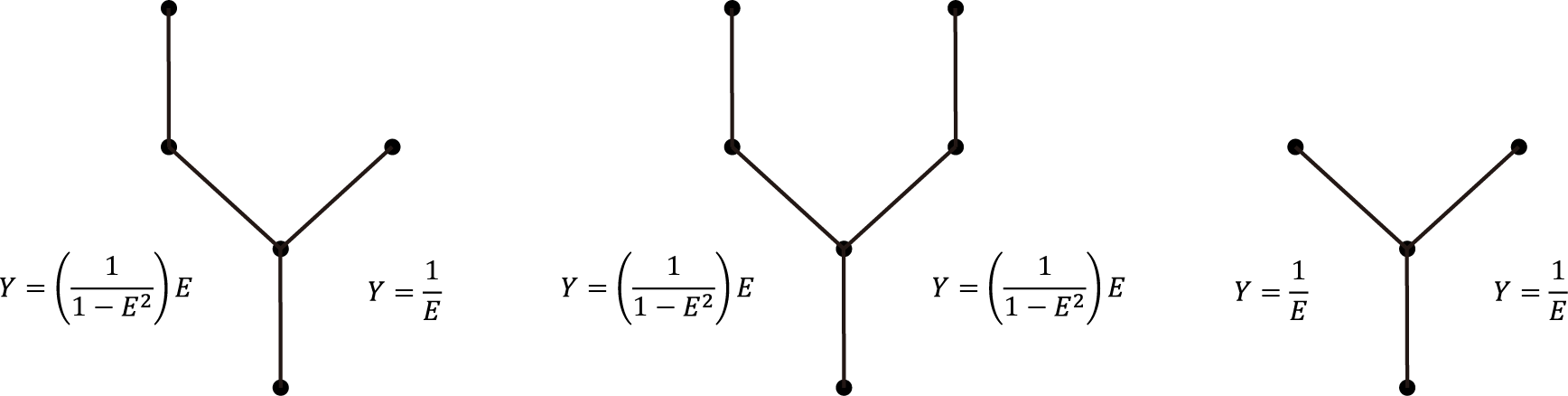}
	\caption{\textbf{Three cases of input leaves.} The fraction $Y$ as output for each input case in this layer is also the input for the branch in the next layer.}
	\label{fig_input}
\end{figure*}

As shown in Fig.~\ref{fig_input}, there are three kinds of input for a branch. We can obtain the output fraction $Y$ by applying Hamiltonian to those sites. Any logic bit input $0$ equals to an input $Y=-\infty$, and any logic input bit $1$ equals to an input $Y=0$ when $E=0$.

The output $Y$ in this layer will be the inputs $Y_1$ or $Y_2$ in the next layer. By iterating the steps in the tree, the final output $Y$ of the tree is the fraction of the site 0 and site root $\frac{\braket{r=0|E}}{\braket{\text{root}|E}}$, which is the reciprocal of the fraction $y(E)=\frac{\braket{\text{root}|E}}{\braket{r=0|E}}$. Therefore, $y(E)=0$ when $Y = -\infty$, and $y(E)=\infty$ when $Y = 0$, which gives the computation result of the NAND Tree problem. 

To prepare an exactly zero energy state, the wave-packet should be infinite long due to the Heisenberg uncertain relationship, which is unfeasible in any experiment platforms. Fortunately, for a superposition of energy states with energy closing to 0, the transmit coefficient $T(E)$ could also be close enough to $T(0)$. Mathematically, it can be described as
\begin{equation}
	|T(E)-T(0)|\leq D|E|, \quad |E|\leq\epsilon
\end{equation} 
where the $\epsilon$ and $D$ depends on the size of the tree. 

\subsubsection{Algorithm with Gaussian wave-packet} 

In this section we apply Gaussian wave-packet as the initial state for the quantum NAND tree. The Gaussian wave-packet is described as
\begin{equation} 
	|\psi_{\text{gs}}(t)\rangle = \sum_{r=-L_{\text{rw}}}^{0}A e^{-\frac{(r-\mu)^{2}}{4\sigma^{2}}}e^{-i\pi r/2}|i\rangle,\label{eq:wp}
\end{equation} 
where $A$ is the normalized coefficient $A=(\frac{1}{2\pi\sigma})^{\frac{1}{4}}$. In principle, the Gaussian wave-packet would spread to the whole space in lattice during the evolution. Here, we need to confine the wave-packet in the range of $-L_{\text{rw}}\leq r\leq0$ as much as possible, which can be realized by choosing $\mu$ and $\sigma$ appropriately. To be specific, by setting $\mu = -\frac{L_{\text{rw}}}{2}$ and $L_{\text{rw}}/2 \geq 6\sigma$, one can ensure that more than $99.7\%$ of the wave-packet is located in the desire region. With these parameters, the energy satisfies

\begin{equation}
	\bra{\psi(0)}H\ket{\psi(0)}\approx0,
\end{equation}
and energy spread (energy variance) satisfies
\begin{equation}
	\bra{\psi(0)}H^{2}\ket{\psi(0)}\approx2(1-e^{-\frac{1}{2\sigma^{2}}}).
\end{equation}
By expanding the exponent part, we find that the energy spread (energy variance) is at the order of $1/\sigma$. In other words, the state is the superposition of a set of eigenstates, whose energy is distributed around $0$ with width $1/\sigma$.

Next, we investigate the evolution of the state
\begin{equation}
	\ket{\psi(t)} = e^{-iHt}\ket{\psi(0)} .
\end{equation}
Firstly, we decompose the state $\ket{\psi(t)}$ into two parts as
\begin{equation}
	\ket{\psi(t)}=\ket{\psi_1(t)}+\ket{\psi_2(t)},
\end{equation}
where
\begin{equation}
	\ket{\psi_1(t)} = \int^{\frac{\pi}{2}+\epsilon}_{\frac{\pi}{2}-\epsilon} \frac{d\theta}{2\pi}e^{-i t E(\theta)}\ket{E(\theta)}\braket{E(\theta)|\psi(0)}.
\end{equation}
For convenience, we set $\theta = \phi + \pi/2$, and then
\begin{equation}
	\braket{E(\theta)|\psi(0)}=A\sum_{r=-L_{\text{rw}}}^{0}  (e^{ir\phi}+(-1)^{r}R^{*}(E)e^{-ir\phi})e^{-\frac{(r-\mu)^2}{4\sigma^{2}}}.
\end{equation}

We separate it into two parts as $A(\phi)+B(\phi)$, represent by $\ket{\psi_{A}(t)}$ and $\ket{\psi_{B}(t)}$ respectively. By using momentum function in probability theory, we obtain
\begin{align} 
	A(\phi)&=2A\sqrt{\pi}\sigma e^{i\phi\mu-\sigma^{2}\phi^{2}},\\
	B(\phi)&=2A\sqrt{\pi}\sigma e^{i(\pi-\phi)\mu-\sigma^{2}(\pi-\phi)^{2}},
\end{align}
where $|B(\phi)|^2$ is close to $0$ by order of $O(\frac{1}{\sigma^2})$. We now have
\begin{equation}
	\ket{\psi_B(t)}=\int^{\epsilon}_{-\epsilon} \frac{d\phi}{2\pi}e^{-2it \sin \phi}R^{*}(E)B(\phi)\ket{E(\phi+\pi/2)},
\end{equation}
and then
\begin{equation}
	\braket{\psi_B(t)|\psi_B(t)} = O(\frac{1}{\sigma^2}).
	\label{24}
\end{equation}
Meanwhile,
\begin{equation}
	\int^{-\epsilon}_{-\pi} \frac{d\phi}{2\pi} |A(\phi)|^{2} + \int_{\epsilon}^{\pi} \frac{d\phi}{2\pi} |A(\phi)|^{2} = \text{erfc}(\sqrt{2}\sigma\epsilon)<\frac{1}{\sigma\epsilon}.
	\label{25}
\end{equation}
Combining Eq.~\ref{24} and Eq.~\ref{25}, we find that
\begin{equation} ||\ket{\psi_1(t)}||\geq ||\ket{\psi_A(t)}||-||\ket{\psi_B(t)}||.\end{equation}
Therefore,
\begin{align}
	||\ket{\psi_1(t)}||&=1-O(\frac{1}{\sigma\epsilon}),\\
	||\ket{\psi_2(t)}||&=O(\frac{1}{\sqrt{\sigma\epsilon}}).
\end{align}
It shows that $\ket{\psi_A]}$ is a good approximation to the true state $\ket{\psi(t)}$. Now we can decompose $\ket{\psi_A(t)}$ into four parts as
\begin{align}
	a_r(t)&=T(0)i^r\int_{-\pi}^{\pi}\frac{d\phi}{2\pi}e^{i(r-2t)\phi}A(\phi),\\
	b_r(t)&=-i^r T(0)\{\int_{-\pi}^{-\epsilon} \frac{d\phi}{2\pi} +\int^{\pi}_{\epsilon} \frac{d\phi}{2\pi} \}e^{i(r-2t)\phi}A(\phi),\\
	c_r(t)&=i^r T(0) \int_{-\epsilon}^{\epsilon}\frac{d\phi}{2\pi}(e^{-2it\sin \phi}-e^{-2it\phi})e^{ir\phi}A(\phi),\\
	d_r(t)&= i^r \int^{\epsilon}_{-\epsilon}\frac{d\phi}{2\pi}(T(E)-T(0))e^{-2it \sin \phi}e^{ir\phi} A(\phi),\\
\end{align}
where
\begin{align} 
	\sum^{\infty}_{r=0}|b_r(t)|^2 &= O(\frac{1}{\sigma\epsilon}),\\
	\sum^{\infty}_{r=0}|c_r(t)|^2 &= O(\sigma^3 \epsilon^7),\\
	\sum^{\infty}_{r=0}|d_r(t)|^2 &= O(D^2\sigma\epsilon^3).
\end{align}
Putting all of these four parts together, we can get that
\begin{equation} 
	\sum_{r>0} |\braket{r|\psi (t)}| ^{2}=|T(0)| ^{2}+ O(\frac{1}{(\sigma\epsilon)^{1/4}}, D\sqrt{\sigma\epsilon^3}).
\end{equation}
Thus, for a proper time $t$ at $r>0$, the state is 
\begin{equation} 
	\braket{r|\psi(t)} = T(0)\braket{r-2t|\psi(0)} 
\end{equation}
with a small correction. In this way, if we choose the projector
\begin{equation} 
	P_{+} = \sum _{r>0} \ket{r}\bra{r},
\end{equation}
for measurement, we have
\begin{align} 
	P_{+} &= \sum_{r>0} |\braket{r|\psi (t)}| ^{2}\\
	&=|T(0)| ^{2}+ O(\frac{1}{(\sigma\epsilon)^{1/4}}, D\sqrt{\sigma\epsilon^3}),
\end{align}
where $t$ depends on the original location and the velocity of the wave-packet. In experiment, we repeat the process and measure the state in different time. In Tab.~\ref{tab_Comparison}, we give the comparison of our result with that in the originally plane-wave proposal.

Moreover, by taking $\sigma = O(L_{\text{rw}})$, $L_{\text{rw}}=\gamma \sqrt{N}$, and choosing $\epsilon = \frac{1}{16\sqrt{N}}$, $D=8\sqrt{N}$, we have error probability as $O(\gamma^{-1/4})$, which is independent with $N$ (see Tab.~\ref{tab_result}). With large enough $\gamma$, one can reduce the error probability as close as 0. Such that, we can just investigate whether the measure output is larger than $50\%$ in our experiment.

\begin{table*}
	\centering
	
	\caption{\textbf{The comparison between the plane wave-packet and Gaussian wave-packet.}}
	
	\begin{tabular}{p{4.6cm}<{\centering} p{4.6cm}<{\centering} p{4.6cm}<{\centering}}
		\hline
		\hline
		Wave-packet & Energy expanding                             & Error rate\\[0.2cm]
		\hline\noalign{\smallskip}
		plane & $O(1/\sqrt{L})$    & $O(1/\sqrt{L_{\text{rw}}\epsilon}, D\sqrt{\frac{\epsilon}{L_{\text{rw}}}},(\frac{\sigma}{L_{\text{rw}}})^{1/4})$    \\[0.2cm]
		Gaussian & $O(1/\sigma)$  & $O(\frac{1}{(\sigma\epsilon)^{1/4}}, D\sqrt{\sigma \epsilon^3})$ \\[0.2cm]
		\hline
		\hline
	\end{tabular}
	\label{tab_Comparison}
\end{table*}

\begin{table*}
	\centering

	\caption{\textbf{The estimate of the transmission coefficient for the Gaussian wave-packet.}}
	\begin{tabular}{p{3.45cm}<{\centering} p{3.45cm}<{\centering} p{3.45cm}<{\centering} p{3.45cm}<{\centering}}
		\hline
		\hline
		NAND result & Fraction(y)                             & Transmit efficient     & Energy \\[0.2cm]
		\hline
		0 & $|y|>\frac{1}{4\sqrt{N}|E|}$ & $|T|<8\sqrt{N}|E|$    & $|E|<\frac{1}{16\sqrt{N}}$\\[0.2cm]
		1 & $|y|<4\sqrt{N}|E|$                 & $|T|<1-8\sqrt{N}|E|$ & $|E|<\frac{1}{16\sqrt{N}}$\\[0.2cm]
		\hline
		\hline
	\end{tabular}
	\label{tab_result}
\end{table*}

\subsection{Theoretical analysis: quantum slide}

In this section we will show how to use non-uniform coupling spin chain, or ``Quantum Slide'', to generalize the Gaussian wave-packet with $\sigma = O(\sqrt{L_{\text{qs}}})$ in a ${L_{\text{qs}}}$ sited spin chain.

In a well known perfect spin transfer scheme~\cite{christandl}, a parabolic coupling model can be used for the  perfect state transfer
$\ket{0...01}=e^{-iH\pi/2}\ket{10....0}$ in a $L$ length spin chain. The Hamiltonian is
\begin{equation} 
	H=-\left[
	\begin{matrix}
		0      & J_{1}      & 0 & \cdots & 0 &0\\
		J_{1}      & 0      & J_{2} & \cdots & 0 &0 \\
		0      & J_{2}      & 0 & \cdots & 0 &0\\
		\vdots & \vdots &\vdots & \ddots & \vdots &\vdots\\
		0 & 0 & 0 & \cdots&0 &J_{L-1}\\
		0  & 0 & 0  & \cdots&J_{L-1} & 0      \\
	\end{matrix}
	\right],
	\label{PST}
\end{equation}
where \begin{equation}
	J_i = \sqrt{i(L-i)}.
\end{equation}
Inspired by this, we propose a way to automatically generate a Gaussian wave-packet with Eq.~\ref{PST}. Firstly, Eq.~\ref{PST} can be diagonalized as
\begin{equation}
	H=S^{-1}K\Lambda K^{-1}S
\end{equation}
by a diagonal matrix $S$ with elements $s_{ii}=\sqrt{\binom{L-1}{i-1}}$ and Kravchuk matrix $K$, whose elements satisfy
\begin{equation}
	K_{ij}^{N-1} = \sum_l (-1)^l \binom{N-1-j}{i-l}\binom{j}{l}.
\end{equation}
The eigenvalues of the Hamiltonian is $\Lambda=\text{diag}(1-L,3-L,....L-3,L-1)$. Then, we can calculate the element
\begin{equation}
	\bra{\psi(0)}e^{iHt}\ket{r} = \bra{\psi(0)}S^{-1}Ke^{i\Lambda t}K^{-1}S\ket{r}.
	\label{45}
\end{equation}
For the Kravchuk matrix,
\begin{equation}
	K^{(L)2}=2^L I.
\end{equation}
So the inverse matrix of the Kravchuk matrix is
\begin{equation}
	K^{-1} = 2^{1-L}K.
\end{equation}
Then Eq.~\ref{45} becomes
\begin{equation}
	\bra{\psi(0)}e^{iHt}\ket{r} = 2^{1-N}e^{-it(N-1)}\sqrt{\binom{L-1}{r-1}}\sum_m^{N-1}K_{m, k-1}(e^{2it})^m.
\end{equation}
Meanwhile, with the property of the Kravchuk matrix,
\begin{equation}
	(1-v)^j(1+v)^{N-j}=\sum_{i=0}^N K_{ij}^{(N)}v^i, j=0,1,2,...,N
\end{equation}
we can obtain the final result as
\begin{equation}
	\braket{r|e^{iHt}|\psi(0)}=\sqrt{\binom{L-1}{r-1}}(\cos t)^{L-r} (\sin t)^{r-1} e^{\frac{i\pi r}{2}},
\end{equation}
where we have used the fact that $(\sin t)^2 + (\cos t)^2 = 1$ and $L-r + r-1 = L-1$.

By taking Binomial-Gaussian approximation, the above wave function can be approximated as  a right-moving Gaussian wave-packet in Eq.~\ref{eq:wp} with parameters
\begin{align}
	\mu &= np  \\
	&= (L-1)(\sin t)^2\label{54}\\
	\sigma^2 &= npq \\
	&= (L-1)(\sin t)^2 (\cos t)^2
	\label{56}
\end{align}
where $p = (\sin t)^2$, $q = (\cos t)^2$.

Instead of the entire parabolic spin chain, one can generate a Gaussian wave-packet with the Quantum Slide, a $L_{\text{qs}}$ sited lattice with the coupling between the $r$th and $(r+1)$th site as
\begin{equation}
	J_{r}=\frac{J}{L_{\text{qs}}}\sqrt{r(2L_{\text{qs}}-r)}.
\end{equation}
The parameters of the generated a Gaussian wave-packet are
\begin{align}
	\mu &=(2L_{\rm qs}-1)/2\label{QSmu}\\
	\sigma^2 &= (2L_{\rm qs}-1)/4.
	\label{QSsig}
\end{align}
Moreover, according to our simulation, such Gaussian wave-packet generated in the Quantum Slide can be transferred to the uniform spin chain (runway) smoothly.

\subsection{Properties of the wave-packet}

As discussed above, with the width of the Gaussian wave-packet increasing, the wave-packet becomes more `plane', the energy of the wave-packet comes closer to 0, and the error rate will also decrease. Due that the runway is finite, there is a nature error when we measure the right side, if the wave-packet width goes beyond the width of the left side of the runway.

\begin{figure*}[t]
	\includegraphics[width=1.3\columnwidth]{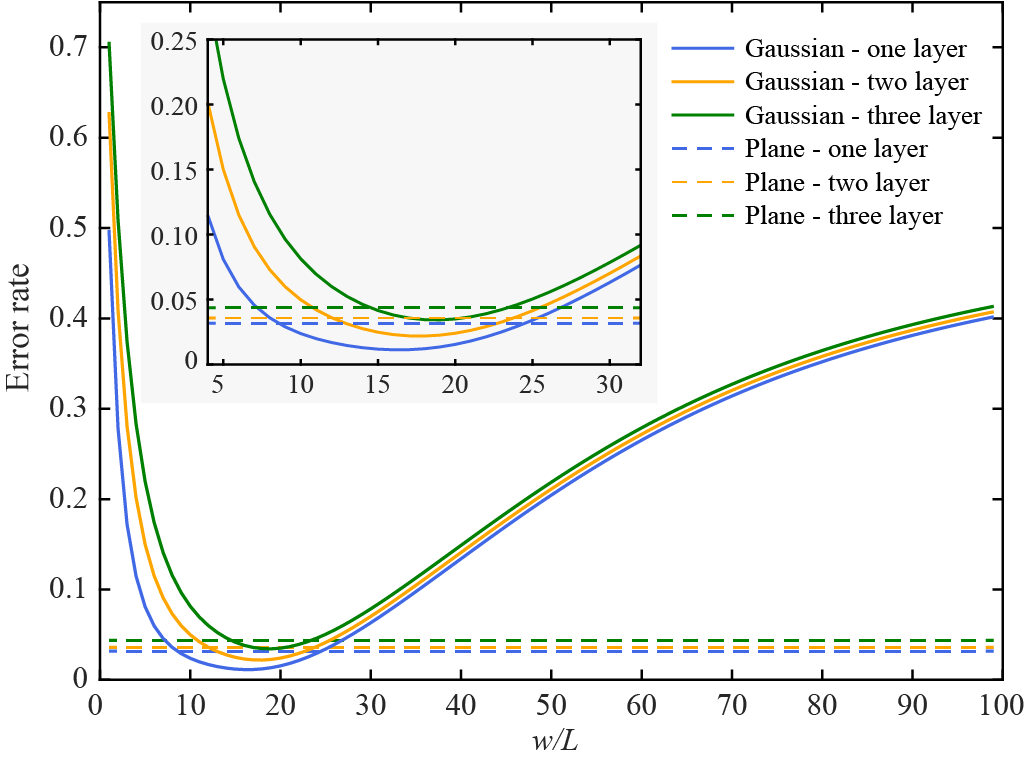}\\
	\caption{\textbf{Error rate with different shape of Gaussian wave-packet and the comparison with plane wave-packet}}
	\label{s-var}
\end{figure*}

\begin{figure*}[t]
	\includegraphics[width=1.3\columnwidth]{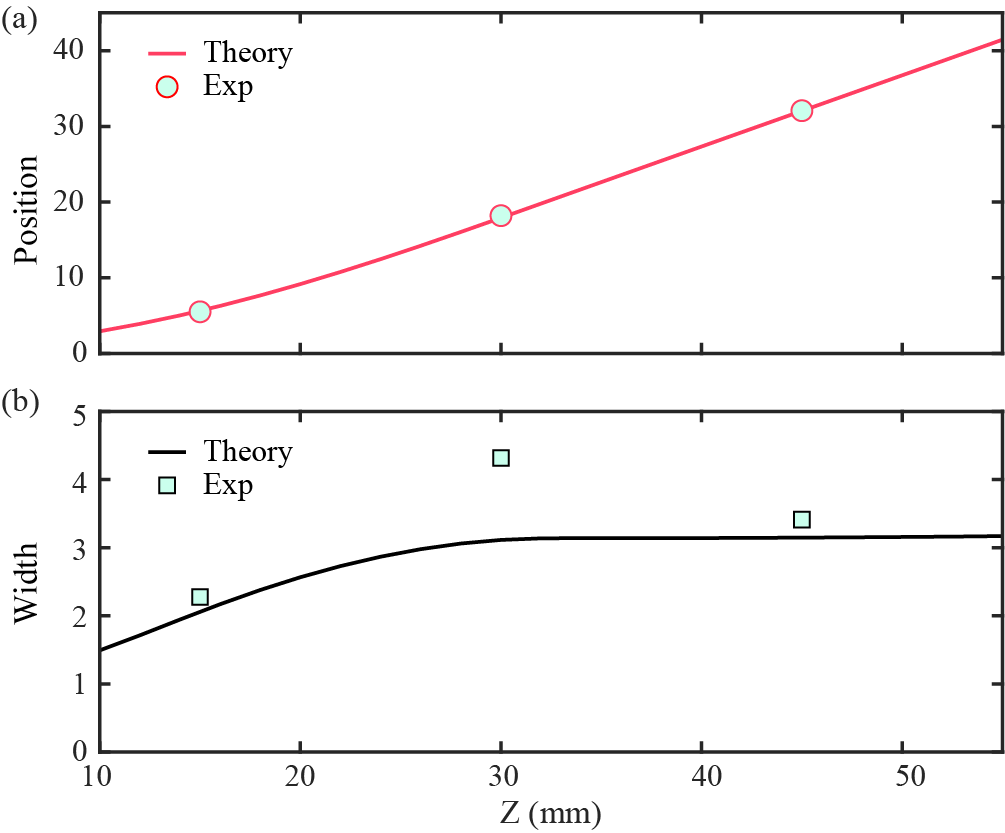}\\
	\caption{\textbf{The theoretical simulation and experimental result of the position and width of the wave-packet.}}
	\label{s3}
\end{figure*}

To find the best parameter of Gaussian wave-packet for this work, we take the a runway with 250 sites in the left side for example. For convenience, here we set the standard deviation as the wave-packet width. As shown in Fig.~\ref{s-var}, when the width of the Gaussian wave-packet is only about a few percent of the runway, the energy is not close to 0 so the error is large. With the width of wave-packet increasing, the energy of the wave-packet reduce and the error rate decreases rapidly. Then the wave-packet width overruns the width of left side of the runway, the error rate increases again. The minimum of the error rate occurs when the standard of the wave-packet is about 1/6 of the allowed length. The physical mechanism behind is that the wave-packet is wide enough to just fit the total space -- about 99.7\% of the body, according to 3-sigma principle when the center of the Gaussian wave-packet is located on the middle of the left runway. In such case, the error rate may even lower than the plane wave situation, suggesting that our model is even better than the original algorithm.

In our experiment, a 20-sited quantum slide is fabricated to generate the Gaussian wave-packet. The standard deviation of the wave-packet can be estimated from Eq.~\ref{QSsig} as $\sigma = \sqrt{(40-1)/4}\approx 3.12$ sites. As shown in Fig.~\ref{s3}, the measured standard deviation of the wave-packet in runway is about 3.24 sites, which is $20\%$ of the left runway with length 16 sites, closing to the best result.

According to Eq.~\ref{54}, we can estimate the right-moving velocity of the wave-packet as
\begin{equation}
	v = \frac{d\mu}{dt} = (2L_{\rm qs}-1)\sin{2t},
\end{equation}
where the $\sin(2t)=1$ due that $L_{\rm qs}$ is set to $L/2$ in the quantum slide. Taking the coupling strength into consideration, we can find
\begin{equation}
	v = J(2L_{\rm qs}-1)/L_{\rm qs}
\end{equation}
In quantum slide, the velocity depends on the strength of coupling, and the wave-packet speeds up in the slide, and then moves in a constant velocity into the uniform coupling runway. In our photonic chip, the coupling strength on the uniform coupling runway is $J=0.48$ mm$^{-1}$, then the theoretical velocity is $v_t=0.936$ site/mm. The simulated result is $v_s=0.941$ site/mm and experimentally measured result is $v_e=0.925\pm0.027$ site/mm.

\subsection{Experimental results}

\begin{figure*}[t]
	\includegraphics[width=1.5\columnwidth]{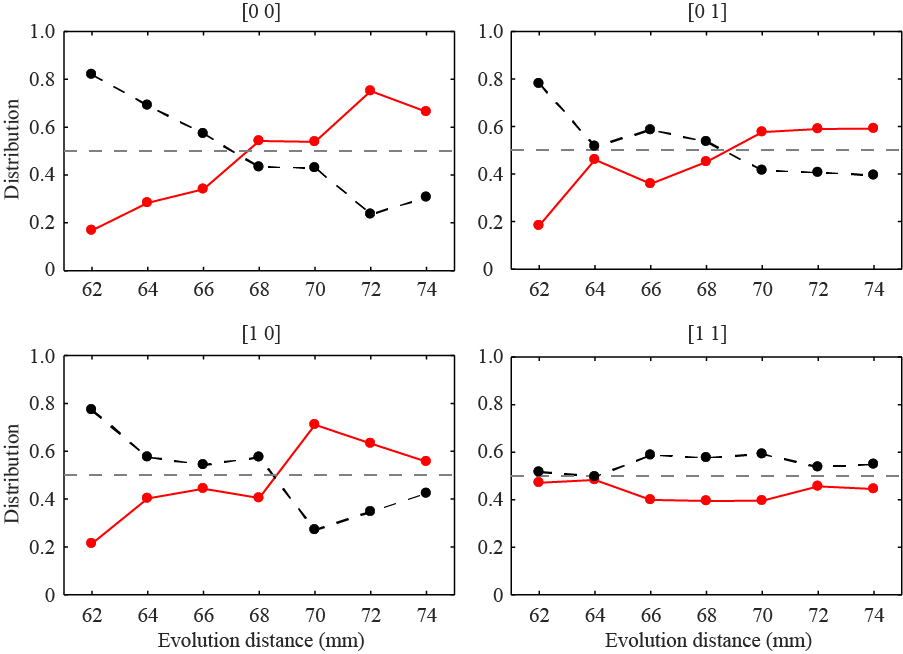}\\
	\caption{\textbf{The dynamical result for the two-bit input NAND tree.}}
	\label{s1}
\end{figure*}

\begin{figure*}[t]
	\includegraphics[width=1.3\columnwidth]{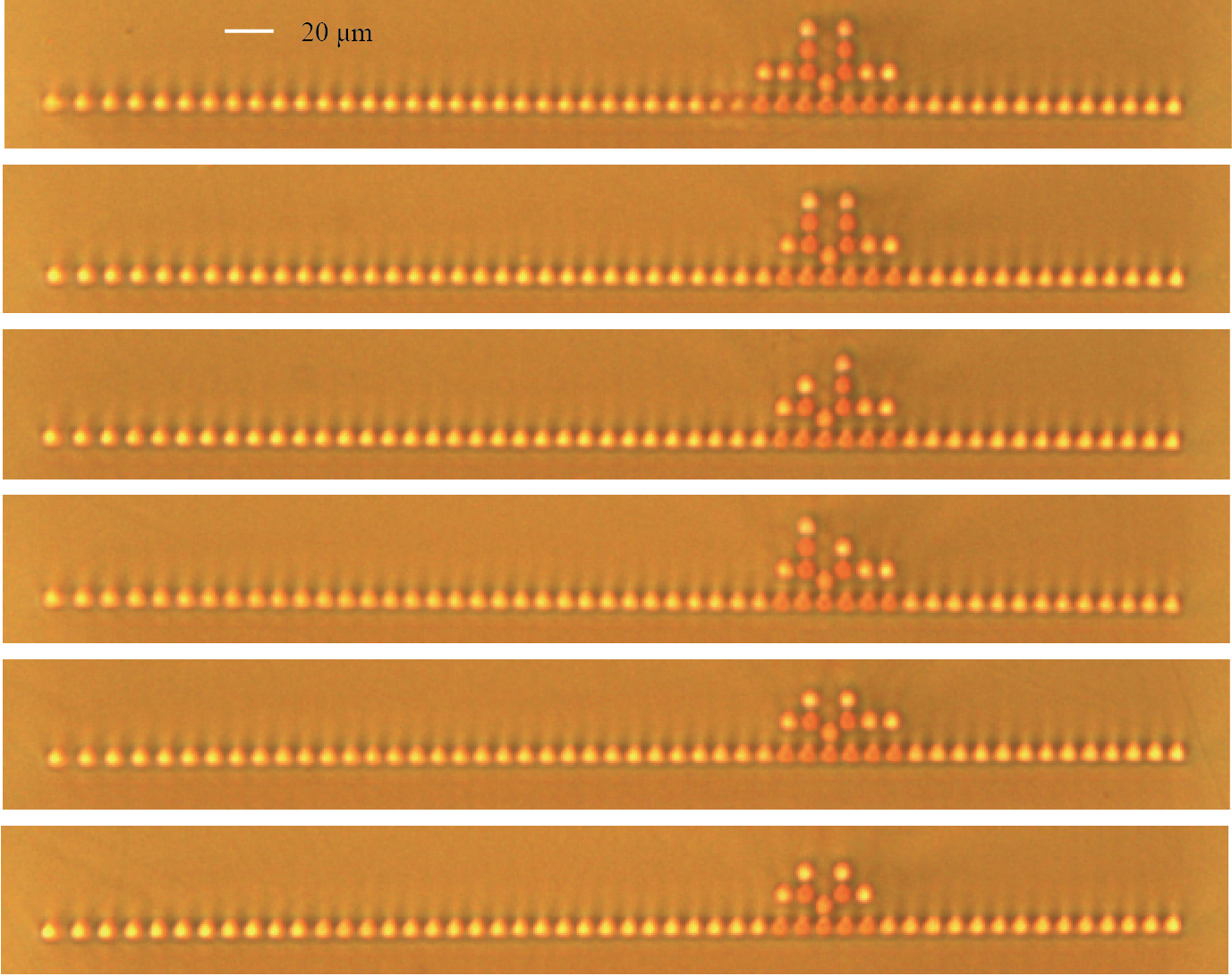}\\
	\caption{\textbf{The cross section micrograms of four-bit NAND tree.}}
	\label{s_facet}
\end{figure*}

\begin{figure*}[t]
	\includegraphics[width=1.6\columnwidth]{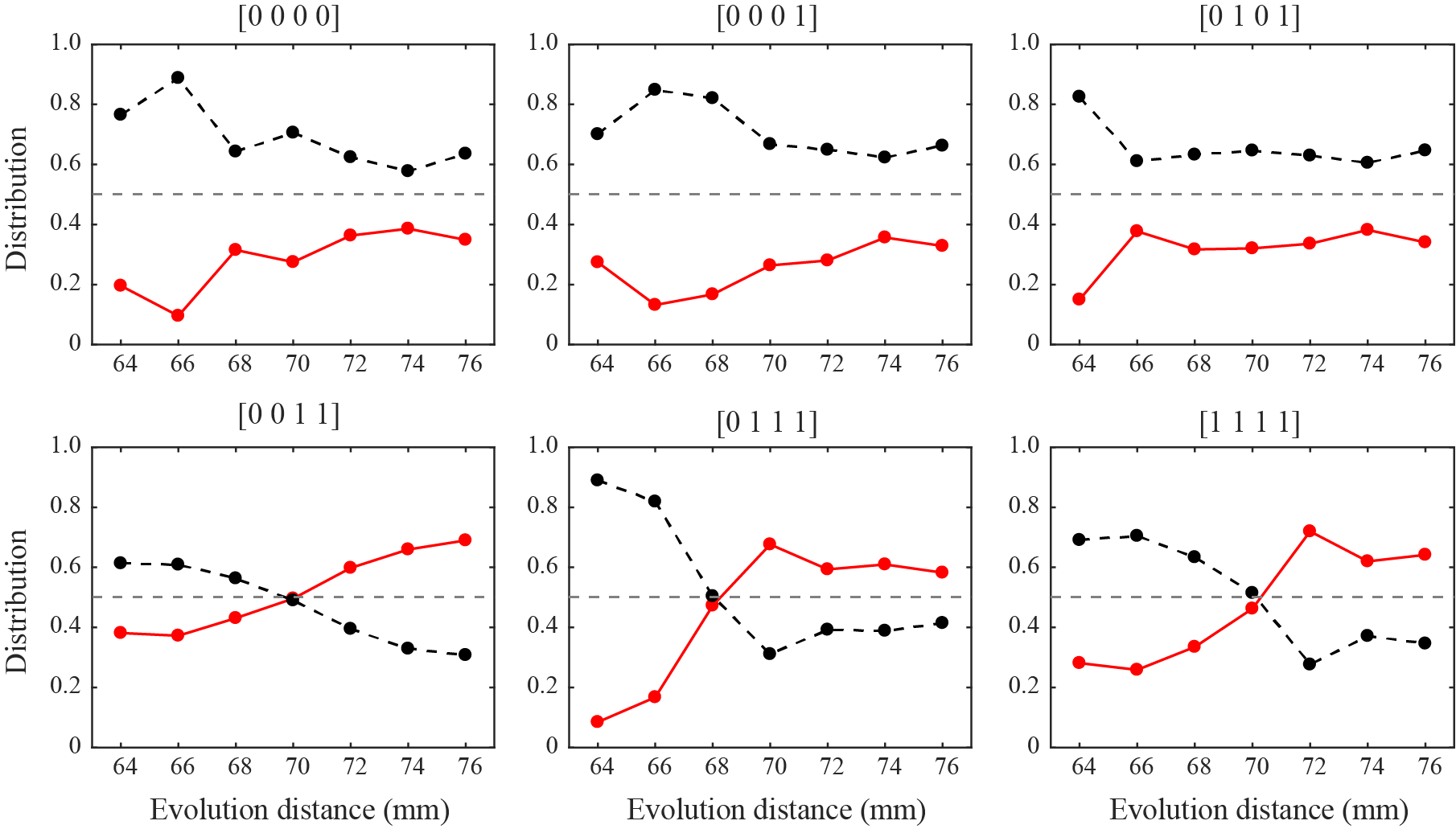}\\
	\caption{\textbf{The dynamical result for the four-bit input NAND tree.}}
	\label{s2}
\end{figure*}

In this section, we show the comprehensive results obtained in our experiment: the dynamical result for the two-bit input NAND tree in Fig.~\ref{s1}, the cross section micrograms of four-bit NAND tree in Fig.~\ref{s_facet}, and the dynamical result for the four-bit input NAND tree in Fig.~\ref{s2}.

\begin{thebibliography}{99}
	
	\bibitem{QW_1} Farhi, E. and Gutmann, S. Quantum computation and decision trees. Phys. Rev. A \textbf{58}, 915-928 (1998).
	\bibitem{QW_1D} Perets, H. B. \textit{et al.} Realization of quantum walks with negligible decoherence in waveguide lattices. Phys. Rev. Lett. \textbf{100}, 170506 (2008).
	\bibitem{QW_co} Peruzzo, A. \textit{et al.} Quantum Walks of Correlated Photons. Science \textbf{329}, 1500-1503 (2010).
	\bibitem{QW_2} Childs, A. M. and Strouse, D. J. Levinson's theorem for graphs. J. Math. Phys. \textbf{52}, 082102 (2011).
	\bibitem{QW_3} Childs, A. M. and Gosset, D. Levinson's theorem for graphs II. J. Math. Phys. \textbf{53}, 102207 (2012).
	\bibitem{photon_sim} Aspuru-Guzik, A. and Walther, P. Photonic quantum simulators. Nat. Phys. \textbf{8}, 285-291 (2012).
	\bibitem{QW_2D} Tang, H. \textit{et al.} Experimental Two-dimensional Quantum Walk on a Photonic Chip. Sci. Adv. \textbf{4}, eaat3174 (2018).
	\bibitem{FH} Tang, H. \textit{et al.} Experimental Quantum Fast Hitting on Hexagonal Graphs. Nat. Photon. \textbf{12}, 754-758 (2018).
		
	\bibitem{BS_1} Broome, M. A.\textit{ et al.} Photonic boson sampling in a tunable circuit. Science \textbf{339}, 794-798 (2013).
	\bibitem{BS_2} Spring, J. B. \textit{et al.} Boson sampling on a photonic chip. Science \textbf{339}, 798-801 (2013).
	\bibitem{BS_3} Tillmann, M. \textit{et al.} Experimental boson sampling. Nat. Photon. \textbf{7}, 540-544 (2013).
	\bibitem{BS_4} Crespi, A. \textit{et al.} Integrated multimode interferometers with arbitrary designs for photonic boson sampling. Nat. Photon. \textbf{7}, 545-549 (2013).
	
	\bibitem{Box} Childs, A. M., Cleve, R., Deotto, E., Farhi, E., Gutmann, S. and Spielman, D. A. In \textit{Proceedings of the thirty-fifth annual ACM symposium on Theory of computing} 59-68 (ACM, San Diego, CA, USA, 2003).
	
	\bibitem{distinctness} Ambainis, A. Quantum Walk Algorithm for Element Distinctness. SIAM J. Comput. \textbf{37}, 210-239 (2007).
	
	\bibitem{QC_1} Deutsch, D. and Jozsa, R. Rapid solution of problems by quantum computation. Proc. R. Soc. A \textbf{439}, 553-558 (1992).
	
	\bibitem{QC_2} Shor, P. W. Polynomial-Time Algorithms for Prime Factorization and Discrete Logarithms on a Quantum Computer. SIAM J. Comput. \textbf{26}, 1484-1509 (1997).
	
	\bibitem{ML} Paparo, G. D., Dunjko, V., Makmal, A., Martin-Delgado, M. A. and Briegel, H. J. Quantum Speedup for Active Learning Agents. Phys. Rev. X \textbf{4}, 031002 (2014).
	
	\bibitem{UCQW0} Childs, A. M. Universal Computation by Quantum Walk. Phys. Rev. Lett. \textbf{102}, 180501 (2009).
	\bibitem{UCQW} Childs, A. M., Gosset, D. and Webb, Z. Universal Computation by Multiparticle Quantum Walk. Science \textbf{339}, 791-794 (2013).
	
	
	\bibitem{Prob1} Saks, M. E. and Wigderson, A. Probabilistic boolean decision trees and the complexity of evaluating game tree. In \textit{Proceedings of the 27th Annual Symposium on Foundations of Computer Science} 29-38 (1986).
	
	\bibitem{Prob2} Santha, M. Probabilistic boolean decision trees and the complexity of evaluating game trees. In \textit{Proceedings of 6th IEEE Structure in Complexity Theory Conference} 180-187 (1991).
	
	
	\bibitem{farhi2007} Farhi, E., Goldstone, J. and Gutmann, S. A quantum algorithm for the Hamiltonian NAND tree. Theory Comput. \textbf{4}, 169-190 (2008). 
	
	
	
	\bibitem{anyNAND} Childs, Andrew M., \textit{et al.} Every NAND formula of size N can be evaluated in time $N^{1/2+ o (1)}$ on a quantum computer. Preprinted in \textit{arXiv:0703015} (2007).
	
	\bibitem{anyANDOR} Ambainis, A. \textit{et al.} Any AND-OR formula of size N can be evaluated in time  $N^{1/2+ o (1)}$ on a quantum computer. SIAM J. Comput. \textbf{39}, 2513-2530 (2010).
	
	\bibitem{mm} Cleve, R., Gavinsky, D., and Yonge-Mallo, D. L. Quantum algorithms for evaluating min-max trees. In \textit{Workshop on Quantum Computation, Communication, and Cryptography} 11-15 (2008). 
	
	\bibitem{games} Reichardt, B. W. Faster quantum algorithm for evaluating game trees. In \textit{Proceedings of the twenty-second annual ACM-SIAM symposium on Discrete algorithms} 546-559 (2011).
	
	
	
	
	\bibitem{discrete_NAND} Ambainis, A. A nearly optimal discrete query quantum algorithm for evaluating NAND formulas. Preprinted in \textit{arXiv:0704.3628} (2007).
	\bibitem{discrete} Childs, A. M., Cleve, R., Jordan, S. P. and Yonge-Mallo, D. Discrete-Query Quantum Algorithm for NAND Trees. Theory Comput. \textbf{5}, 119-123 (2009).
	
	\bibitem{molecular_NAND} Jensen, P. W. K., Jin, C., Dallaire-Demers, P.-L., Aspuru-Guzik, A. and Solomon, G. C. Molecular realization of a quantum NAND tree. Quantum Sci. Technol. \textbf{4}, 015013 (2018).
	
	\bibitem{fabri_1} Davis, K. M., Miura, K., Sugimoto, N. and Hirao, K. Writing waveguides in glass with a femtosecond laser. Opt. Lett. \textbf{21}, 1729-1731 (1996).
	\bibitem{fabri_2} Szameit, A., Dreisow, F., Pertsch, T., Nolte, S. and T{\"u}nnermann, A. Control of directional evanescent coupling in fs laser written waveguides. Opt. Express \textbf{15}, 1579-1587 (2007).
	\bibitem{fabri_3} Chaboyer, Z., Meany, T., Helt, L. G., Withford, M. J. and Steel, M. J. Tunable quantum interference in a 3D integrated circuit. Sci. Rep. \textbf{5}, 9601 (2015).
	\bibitem{PIT_Gap} Wang, Y. \textit{et al.} Parity-Induced Thermalization Gap in Disordered Ring Lattices. Phys. Rev. Lett. \textbf{122}, 013903 (2019).
	
	\bibitem{PST} Christandl, M., Datta, N., Ekert, A. and Landahl, A. J. Perfect State Transfer in Quantum Spin Networks. Phys. Rev. Lett. \textbf{92}, 187902 (2004).

\end{thebibliography}

\begin{thebibliography}{80}
	\bibitem{farhi2007} Farhi, E., Goldstone, J. and Gutmann, S. A quantum algorithm for the Hamiltonian NAND tree. Theory Comput. \textbf{4}, 169-190 (2008). 
	\bibitem{christandl} Christandl M. \textit{et al.} Perfect State Transfer in Quantum Spin Networks. Phys. Rev. Lett. \textbf{92}, 187902 (2004).
\end{thebibliography}
\end{document}